# Effects of Repetitive SSVEPs on EEG Complexity using Multiscale Inherent Fuzzy Entropy

Zehong Cao[a,b*], Tim Chen[a,b,], Weiping Ding[b,d], Yu-Kai Wang[a,b,c], Farookh Khadeer Hussain[a,b], Adel Al-Jumaily[b], Chin-Teng Lin[a,b,c]

[1] Centre for Artificial Intelligence, University of Technology Sydney, Australia
[2] Faculty of Engineering and Information Technology, University of Technology Sydney, Australia
[3] Brain Research Centre, National Chiao Tung University, Taiwan
[4] School of Computer Science and Technology, Nantong University, China

| AETICLE INFORMATION | ABSTRACT |
|---|---|
|  | Multiscale inherent fuzzy entropy is an objective measurement of electroencephalography (EEG) complexity, reflecting the habituation of brain systems. Entropy dynamics are generally believed to reflect the ability of the brain to adapt to a visual stimulus environment. In this study, we explored repetitive steady-state visual evoked potential (SSVEP)-based EEG complexity by assessing multiscale inherent fuzzy entropy with relative measurements. We used a wearable EEG device with Oz and Fpz electrodes to collect EEG signals from 40 participants under the following three conditions: a resting state (closed-eyes (CE) and open-eyes (OE) stimulation with five 15-Hz CE SSVEPs and stimulation with five 20-Hz OE SSVEPs. We noted monotonic enhancement of occipital EEG relative complexity with increasing stimulus times in CE and OE conditions. The occipital EEG relative complexity was significantly higher for the fifth SSVEP than for the first SSEVP (FDR-adjusted $p < 0.05$). Similarly, the prefrontal EEG relative complexity tended to be significantly higher in the OE condition compared to that in the CE condition (FDR-adjusted $p < 0.05$). The results also indicate that multiscale inherent fuzzy entropy is superior to other competing multiscale-based entropy methods. In conclusion, EEG relative complexity increases with stimulus times, a finding that reflects the strong habituation of brain systems. These results suggest that multiscale inherent fuzzy entropy is an EEG pattern with which |

* Correspondence and reprint requests to Dr. Zehong Cao (Zehong.Cao@uts.edu.au), School of Software, Faculty of Engineering and Information Technology, University of Technology Sydney, 15 Broadway, Ultimo NSW 2007, Australia.

**1. Introduction**

A visual stimulus environment has been proposed to be a powerful application by which the activation and sensitization of brain activities can be characterized. The brain visual system is intimately connected with the nervous and endocrine systems by way of the eyes, through which the light receptors of the retinas send electrical messages to the cortical electroencephalography (EEG). The effects of these messages are not only limited to the occipital region but are also seen in the frontal regions [1]. Clinical studies have reported that visual stimuli are effective in inducing relaxation, reducing stress, and relieving insomnia [2, 3].

In particular, steady-state visual evoked potentials (SSVEPs) are natural responses to visual stimuli at specific frequencies. When the retina is excited by a visual stimulus ranging from 3.5 Hz to 75 Hz, the brain generates electrical activity in the visual cortex [4]. The frequency of this activity is a multiple of or the same as the frequency of the visual stimulus. Previous studies have demonstrated that EEG signals over the visual cortex are natural responses to flickering stimuli between 5 and 27 Hz, with the strongest response at frequencies approximately 15 or 20 Hz [5, 6], which can be used to develop brain–computer interface applications [7, 8]. Based on the behavioral characteristics of brain electrical activity, a response decrement resulting from repeated visual stimulation is defined as habituation, suggesting robustness of the brain system [9, 10]. Habituation is generally acknowledged to be characterized by reductions in negative amplitudes in response to repetitive visual evoked potentials [11]. Thus, habituation is generally considered a useful index for investigating neuronal substrates of behavior, mechanisms of learning processes, and treatments of information in the central nervous system [12].

However, the habituation index is generally acknowledged as a linear (amplitude) description of brain dynamics and is thus less likely to provide information about nonlinear (complex) brain dynamics. Recent improvements in the understanding of brain dynamics are attributable to entropy analysis approaches, which assess how complexity provides information about a wide range of physiological systems [13]. Entropy is generally an objective measure of how the complexity of physiological signals represents the robustness of brain systems [14]. Different entropy analysis approaches, such as approximate entropy [15], dispersion entropy [16], original and refined sample entropy [17, 18], and original and refined fuzzy entropy [19, 20] were developed to measure physiological signals.

An integrated brain system is often multiscaled and interacts with faster and slower processes, depending on the scale at which it examines the bio-signal. Early entropy measures do not take into account the multiple time scales in physical systems which quantify the irregularity of a signal. In the 2000s, the multiscale entropy approach was proposed to represent the complexity of a signal [21]. Multiscale entropy relies on the computation of the sample entropy and coarse-grained time series that represent the system dynamics on different scales. In 2009, Valencia et al. [16] proposed refined multiscale entropy to remove the fast temporal scales and used a coarse graining that prevents influence of the reduced variance on the complexity evaluation. Since vector similarity is defined by the hard and sensitive boundary of the Heaviside function in approximate entropy and sample entropy, recent studies have proposed fuzzy entropy and its multiscale version to measure the matching degree of two vectors by using the fuzzy membership function to reduce the sensitivity [19, 20].

Since EEG is composed of nonlinear signals, intrinsic modes extracted from empirical mode decomposition can benefit by eliminating noise/trends in EEG signals, which can improve EEG complexity evaluation. Previous multiscale-based entropy methods did not take into account the inherent measurement, so we recently developed a multiscale inherent fuzzy entropy algorithm [22] that has the robustness to operate noise and nonlinearity signals and is capable of operating EEG signals across a range of temporal (time) scales.

A growing body of literature has reported that visual stimulation attenuates symptoms of physical and mental discomfort [3] and brain dysfunction. For example, visual stimuli have been reported to attenuate migraine disorder [23] and Alzheimer's disease [13], as they lead to decreases in entropy in affected individuals compared to that in healthy individuals. As humans decrease or cease responses to a stimulus after repeated presentations [24], our hypothesis is that the brain system demonstrates stronger robustness during repetitive stimulation, a phenomenon accompanied by increases in EEG complexity.

In the past few years, some studies have investigated SSVEP stimuli changes of EEG power spectra and explored EEG multiscale-based entropy methods in the resting state condition. However, to the best of our knowledge, no studies have investigated brain complexity in a visual stimulus environment. In our study, we combined the two separated perspectives (SSVEP stimuli and multiscale-based entropy methods) and then proposed a relative multiscale inherent fuzzy entropy approach with SSVEP stimuli that is considered as a new study to contribute neuro-engineering, neuroscience and clinical applications. More specifically, we designed a visual stimulus experiment using repetitive SSVEPs to investigate multiscale inherent fuzzy entropy with relative measurements and then assessed the effects of SSVEP stimuli on EEG complexity in open-eyes (OE) and closed-eyes (CE) conditions. Our study may provide an EEG pattern with which complex brain dynamics can be measured in a visual stimulus environment.

## 2. Materials and Methods
### 2.1 Wearable EEG

EEG signals were recorded at a sampling rate of 500 Hz by a "Mindo" EEG device (Brain Rhythm Inc., Zhubei District, Hsinchu, Taiwan), which is a wearable EEG device with dry sensors [25]. Each dry-contact electrode was designed to include a probe head, plunger, spring, and barrel. The probes were inserted into a flexible substrate via an established injection molding procedure using a one-time forming process. These dry electrodes are more convenient for measuring EEG signals than conventional wet electrodes and are preferred because they avoid conductive gel use and skin preparation procedures while achieving a signal quality comparable to that of wet electrodes. In this study, two dry-contact electrodes (Fpz and Oz) were placed according to the extended International 10–20 system, and two extra channels (A1 and A2) were used as reference channels.

### 2.2 Experimental Paradigm

The EEG experiment was performed in a static and lightless room at National Chiao Tung University, Taiwan. To avoid light source interference, we turned the fluorescent lamps off during the experimental procedure. The display was placed approximately 20 cm in front of the participants' eyes, and the repetitive visual stimulus was presented on the screen in the form of alternating graphical patterns.

As shown in Fig. 1, this experiment consisted of the following three sessions: a resting (CE and OE) session, a CE visual stimulus session and an OE visual stimulus session. The resting session comprised three epochs each in which the eyes were open and closed for 1 min. To test the effects of a strong SSVEP ranging from 13–25 Hz [26], we administered 5-10-s 15-Hz SSVEPs at 10-s intervals during the CE visual stimulus session. After a 2-min rest period, we completed the OE visual stimulus session, which consisted of 5-10-s 20-Hz SSVEPs administered at 10-s intervals. Of note, we chose different flicker frequencies between the CE and OE sessions to assess the strong effects of SSVEP stimuli.

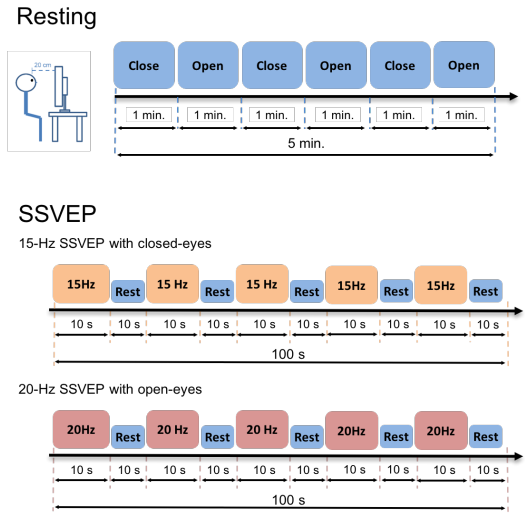

Figure 1 Experimental paradigm in resting and SSVEP conditions

## 2.3 Participants

Forty participants, including 15 males and 25 females aged 28.1 ± 7.2 years, were enrolled in the study. These participants were students and staff from National Chiao Tung University who did not have prior personal or family histories of disease. All participants included in the study had normal or corrected-to-normal vision and no visual impairments. Furthermore, each participant confirmed that he or she had no history of adverse reactions to flashing lights. The Institutional Review Board of National Chiao Tung University approved the study. Informed consent was obtained from all subjects before they entered the study.

## 2.4 EEG Preprocessing

The original EEG was reviewed by experienced EEG specialists. EEG activity was sampled at 500 Hz, which served as the recording sampling rate. The raw EEG signals were down-sampled to 250 Hz and then filtered through 1-Hz high-pass and 30-Hz low-pass finite impulse-response filters. Segments contaminated with nonphysiological artifacts, including movement artifacts, electrode pops, sweating artifacts, and 60-Hz noises, were marked and discarded. Furthermore, the filtered EEG signals were inspected again using the automatic continuous rejection function to remove noisy signals. The EEG signals for the resting and SSVEP trials were extracted from the Oz and Fpz electrodes, respectively. Finally, the epoch EEG data were estimated by the multiscale inherent fuzzy entropy algorithm. All EEG data were analyzed by the EEGLAB (http://sccn.ucsd.edu/eeglab/) with Matlab Software (Mathworks, Inc.), which is an open-source toolbox for electrophysiological signal processing.

## 2.5 EEG Complexity: Multiscale Inherent Fuzzy Entropy with Relative Measurements

EEG complexity is estimated by the multiscale inherent fuzzy entropy algorithm [22] with relative measurements, which is divided into the following four parts: 2.5.1, an empirical mode decomposition (EMD) technique for the detrending process; 2.5.2, a multiscale procedure; 2.5.3, the FuzzyEn algorithm for irregularity evaluation; and 2.5.4, relative complexity (RC) estimation by differences in the multiscale inherent fuzzy entropy between the baseline and stimulus sessions. These four parts are described in Fig. 2 and the following subsections.

2.5.1 EMD technique for the de-trending process (Fig. 2-A)

EMD is an adaptive time-frequency data analysis method [27]. EMD has proven to be quite versatile in a broad range of applications for extracting signals from data generated in noisy and nonlinear processes. EMD performs operations that partition a series into intrinsic mode functions (IMFs) without leaving the time domain.

We applied the EMD technique to decompose the preprocessed signal $x(t)$ into several IMFs and reconstructed the signal $\hat{x}(t)$.

In the initial step, extrema of the signal $s(t)$ are found, corresponding to $E_{\text{minima}}$ and $E_{\text{maxima}}$. Then, the regions between $E_{\text{minima}}$ and $E_{\text{maxima}}$ are interpolated, yielding an envelope with $en_{min}(t)$ and $en_{\max}(t)$.

First, we compute the mean:
$$M(t) = (en_{min}(t) + en_{max}(t))/2 \quad (1)$$

Second, we extract the candidate of inherent functions:
$$Ca(t) = s(t) - M(t) \quad (2)$$
Third, we confirm $Ca(t)$, belonging to an IMF. If $Ca(t)$ satisfies the constraint conditions, $Ca(t)$ is saved, and the residue is computed:
$$res(t) = s(t) - \sum_{i=1}^{t} Ca(t) \quad (3)$$
Next, we solve $t = t + 1$ and treat $res(t + 1)$ as input data. Otherwise, we treat $Ca(t + 1)$ as input data. Iterations are performed on the residual $res(t)$ and continued until the final residue $r$ satisfies the stopping criterion.

Finally, the components of the IMFs with surviving high trends are automatically removed by a trend filter. The signal $\hat{s}(t)$ is reconstructed by the cumulative sum of the remaining IMFs:
$$\hat{s}(t) = \sum_{i=a}^{i=b} Ca(t) \quad (4)$$
The parameter $i$ is the order number of the components from the IMFs, and parameters $b$ and $a$ are the upper and lower boundaries of the selected components, respectively. In our analysis, the parameters $b$ and $a$ were chosen as 10 and 5, respectively.

2.5.2 Multiscale procedure (Fig. 2-B)

The multiscale procedure involves coarse graining the signals into different time scales. For a given time series, multiple coarse-grained time series are constructed by averaging the data points within nonoverlapping windows of increasing length, and the $\tau$ element of the coarse-grained time series $y_j^{(\tau)}$ is calculated according to the following equation:
$$y_j^{(\tau)} = \frac{1}{\tau}\sum_{i=(j-1)\tau+1}^{j\tau} x_i \quad (5)$$
where $\tau$ represents the scale factor and $1 \leq j \leq N/\tau$. The length of each coarse-grained time series is $N/\tau$. For scale 1, the coarse-grained time series is simply the original time series. The appropriate scales (e.g., $\tau = 10$) must be chosen before the FuzzyEn algorithm is calculated. In our analysis, we choose the $\tau$ range from 1 to 20.

2.5.3 Fuzzy entropy algorithm for irregularity evaluation (Fig. 2-C)

The initial step is to normalize the EEG signal using the Z-score measurement. The EEG signal $y(t)$ subtracts the mean prior to dividing by the standard deviation (SD). The normalized EEG signal is marked as $\hat{y}(t)$.

First, considering the $N$ sample time series $\{\hat{y}(i): 1 \leq i \leq N\}$, given $m$, $n$, and $r$, a vector set sequence $\{Y_i^m, i = 1, \ldots, N - m + 1\}$ is calculated, and the baseline is removed:

$$Y_i^m = \begin{Bmatrix} \hat{y}(i), \hat{y}(i+1)\ldots, \\ \hat{y}(i+m-1) \end{Bmatrix} - m^{-1}\sum_{j=0}^{m-1} \hat{y}(i+j)$$

(6)

where $1 \leq i \leq N - m + 1$, and $X_i^m$ presents $m$ consecutive $u$ values, beginning with the $i$th point.

Second, given a vector $Y_i^m$, the similarity degree $D_{ij}^m$ between $Y_i^m$ and $Y_j^m$ is defined by the fuzzy membership function:

$$D_{ij}^m = fu(d_{ij}^m, n, r) = exp\left(-\frac{\left(d_{ij}^m\right)^n}{r}\right)$$

(7)

where the fuzzy membership function $fu$ is an exponential function and $d_{ij}^m$ is the maximum absolute difference between the corresponding scalar components of $Y_i^m$ and $Y_j^m$.

Then, the function $\varphi^m$ is constructed. Similarly, for $m + 1$, the above steps are repeated and denoted $\varphi^{m+1}(n, r)$.

$$\varphi^m(n,r) = (N-m)^{-1}\sum_{i=1}^{N-m}((N-m-1)^{-1}\sum_{j=1, j\neq i}^{N-m} D_{ij}^m) \quad (8)$$

Finally, the $entropy\,(m, n, r, N)$ parameter of the sequence $\{\hat{x}(i): 1 \leq i \leq N\}$ is defined as the negative natural logarithm of the deviation of $\varphi^m$ from $\varphi^{m+1}$:

$$entropy(m, n, r, N) = \ln \varphi^m(n, r) - \ln \varphi^{m+1}(n, r) \quad (9)$$

For the parameter choices of fuzzy entropy, the first parameter $m$ is the length of the sequences to be compared, such as in approximate entropy and sample entropy. The other two parameters, $r$ and $n$, determine the width and the gradient of the boundary of the fuzzy membership function, respectively. The last parameter, $N$, represents a sample time series of EEG signals. In our analysis, we set the parameters as $m = 2, r = 0.15 \times SD,$ and $n = 2$.

2.5.4 Relative complexity (Fig. 2-D)

To minimize individual differences, we introduced relative measurements of EEG multiscale inherent fuzzy entropy to assess RC. First, the inherent fuzzy entropy at baseline (resting) and during stimulus sessions is calculated. The two sessions are simplified to $C_{baseline}$ and $C_{SSVEP}$, respectively.

Next, we calculated the variation in the multiscale inherent fuzzy entropy between the baseline (resting) condition and the stimulus condition during which SSVEPs were induced in five stimulus trials. This function is termed the inherent fuzzy entropy with relative measurements, which is expressed as:

$$RC_i = C_{SSVEP(i)} - C_{baseline} \quad (10)$$

where $i$ is the stimulus time ($i = 1, 2, 3, 4,$ and $5$).

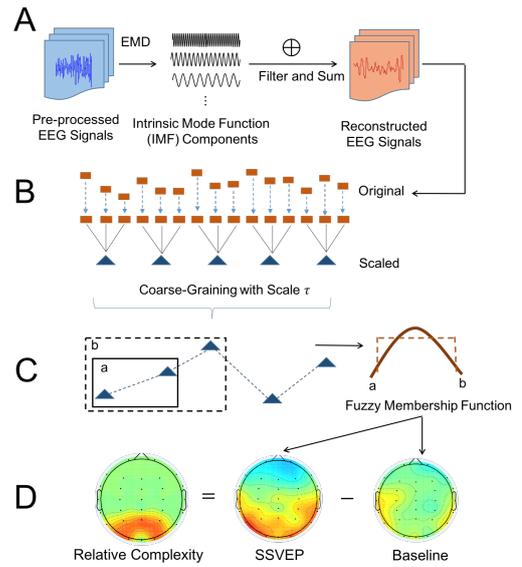

Figure 2 Multiscale inherent fuzzy entropy algorithm

## 2.6 Statistical Analysis

First, we applied the Kolmogorov-Smirnov test (K-S test) to assess the normality of samples (entropy variables). To determine the independence of the different entropy variables in repetitive visual stimulation, we performed the paired *t*-test for each entropy scale to compare EEG complexity between the different conditions (CE vs. OE). Furthermore, one-way ANOVA followed by Tukey's post-hoc test was used to test pairwise comparisons pertaining to the five SSVEP stimuli (first SSVEP vs. fifth SSVEP). The false discovery rate (FDR) correction was used to control for multiple comparisons. All statistical tests were two-tailed, and statistical significance was set at $p < 0.05$.

## 3. Results

The EEG RC, which was assessed with multiscale inherent fuzzy entropy with relative measurements over different time scales ranging from 1 to 20 in CE and OE conditions using Oz and Fpz electrodes, is shown in Figs. 3 and 4. Regarding the effects of repetitive SSVEP stimuli, we noted monotonic enhancement of occipital RC with increasing stimulus times in CE and OE conditions. We also noted that prefrontal RC tended to increase in the OE condition. Additionally, we compared the difference between entropy variables and a standard normal distribution, and the results showed that *h* was 0, which indicates that the K-S test failed to reject the null hypothesis at the default 5% significance level.

### 3.1 Occipital EEG RC

Occipital RC, which was measured with multiscale inherent fuzzy entropy over different time scales ranging from 1 to 20 in CE and OE conditions, is shown in Fig. 3. As shown in Fig. 3-A, occipital RC was monotonically enhanced with increasing stimulus times in both conditions, and the RC value increased with the increasing time scale in the two conditions. The paired *t*-tests revealed that the EEG RC in the OE condition was significantly higher than that in the CE condition in most time scales ($p < 0.05$).

As shown in Fig. 3-B, Tukey's post-hoc test revealed that EEG RC increased significantly during the fifth SSVEP stimulus compared to that during the first SSVEP stimulus in most time scales in the CE condition (FDR-adjusted $p < 0.05$). Similarly, the EEG RC also tended to significantly increase from the first to the fifth SSVEP stimulus in most time scales in the OE condition (FDR-adjusted $p < 0.05$).

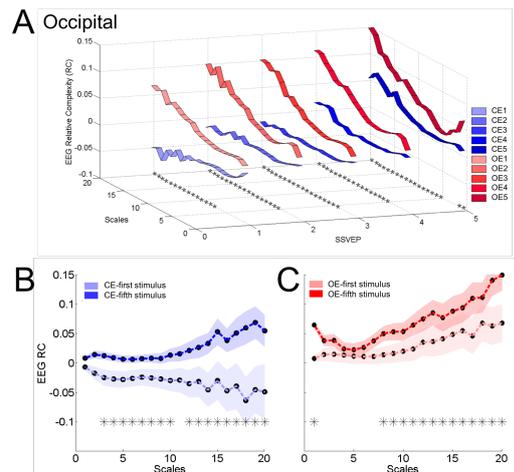

Figure 3 The trends of EEG RC in the occipital area. (A) The changes in the EEG RC during the five SSVEP stimuli (mark 1, 2, 3, 4, 5) over time scales from 1 to 20 in CE and OE conditions. (B) Comparison of EEG RC between the first and fifth stimuli over time scales from 1 to 20 in the CE condition. (C) Comparison of EEG RC between the first and fifth stimuli over time scales from 1 to 20 in the OE condition. Of note, the blue and red traces represent the mean ± SD of the EEG RC of the CE and OE conditions, respectively. The black asterisk in Fig. 3-A denotes a significant difference in EEG RC between the CE and OE conditions (FDR-adjusted $p < 0.05$). The black asterisks in Fig. 3-B and Fig. 3-C denote that EEG RC was significantly increased during the fifth stimulus compared to that during the first stimulus (FDR-adjusted $p < 0.05$).

### 3.2 Prefrontal EEG Complexity

The prefrontal RC, which was measured with multiscale inherent fuzzy entropy over different time scales ranging from 1 to 20 in CE and OE conditions, is shown in Fig. 4. As shown in Fig. 4-A, monotonic enhancements of prefrontal RC occurred with increasing stimulus times only in the OE condition, but the RC values increased with the increasing time scale in both conditions. The EEG RC of the OE condition was significantly lower than that of the CE condition in certain time scales during the first three SSVEP stimuli ($p < 0.05$).

There was no significant difference in the EEG RC between the first and fifth SSVEP stimuli in the CE condition (Fig. 4-B). The EEG RC of the fifth SSVEP stimulus was significantly higher than that of the first SSVEP stimulus in most time scales in the OE condition (FDR-adjusted $p < 0.05$; Fig. 4-C).

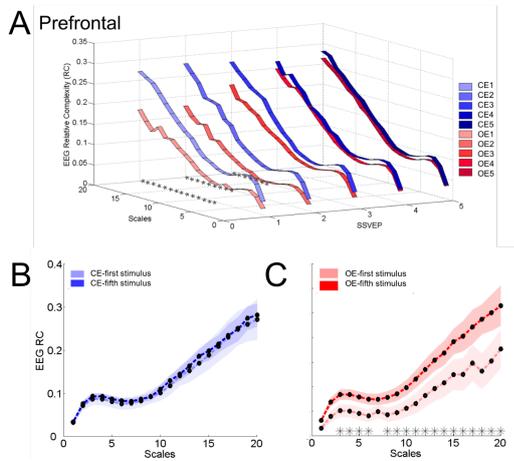

Figure 4 The trends of EEG RC in the prefrontal area. (A) The changes in the EEG RC during the five SSVEP stimuli (mark 1, 2, 3, 4, 5) over time scales from 1 to 20 in CE and OE conditions. (B) Comparison of the EEG RC between the first and fifth stimuli over time scales from 1 to 20 in the CE condition. (C) Comparison of the EEG RC between the first and fifth stimuli over time scales from 1 to 20 in the OE condition. Of note, the blue and red traces represent the mean ± SD of the EEG RC of the CE and OE conditions, respectively. The black asterisk in Fig. 4-A denotes that the EEG RC differed significantly between the CE and OE conditions ($p < 0.05$). The black asterisk in Fig. 4-C denotes that the EEG RC was significantly increased during the fifth stimulus compared to that during the first stimulus (FDR-adjusted $p < 0.05$).

## 3.3 Performance of Competing Multiscale-Based Entropy Methods

In this section, we used the same EEG processing steps but estimated entropy using the competing entropy algorithms. We selected six competing multiscale-based entropy methods for accessing EEG RC, which included multiscale dispersion entropy, multiscale approximate entropy, multiscale sample entropy, multiscale fuzzy entropy, refined multiscale sample entropy, and refined multiscale fuzzy entropy. The parameter values of the competing multiscale-based entropy methods were the same as those of the relative multiscale inherent fuzzy entropy method. In Fig. 5, we could not distinguish occipital EEG entropy between the first and fifth stimuli sessions using competing multiscale-based entropy methods. Similarly, there was no significant difference in frontal EEG entropy between the first and fifth stimuli sessions (Fig. 6). These findings may indicate that the relative multiscale inherent fuzzy entropy algorithm is superior to the other competing multiscale-based entropy methods for the application of SSVEP-based experiments.

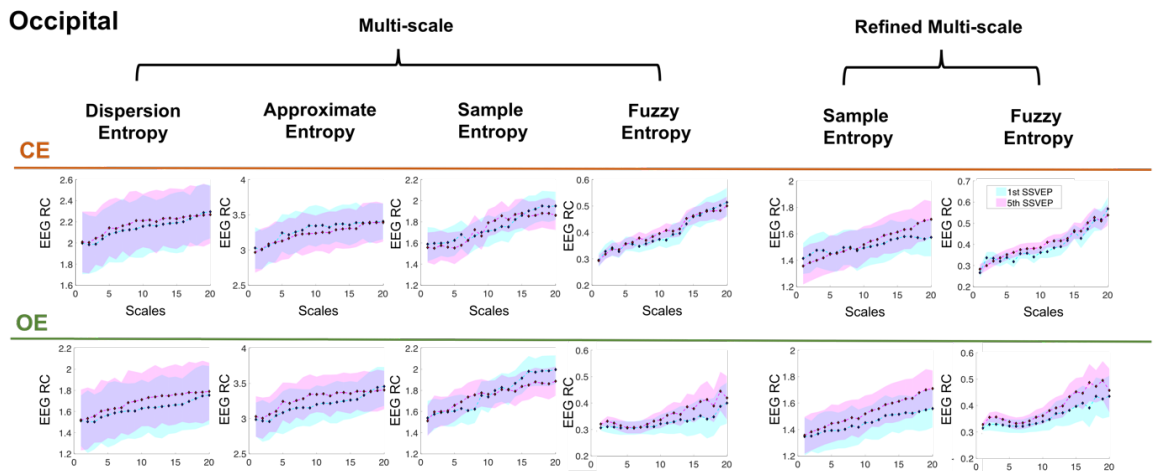

Figure 5 The trends of the EEG RC in the occipital area. Comparison of the EEG RC between the first and fifth stimuli over time scales from 1 to 20 in the CE or OE condition. Of note, the cyan and red traces represent the mean ± SD of the EEG RC in the first and fifth stimuli session, respectively. The purple trace represents the overlap area of the SD of the EEG RC between the first and fifth stimuli session.

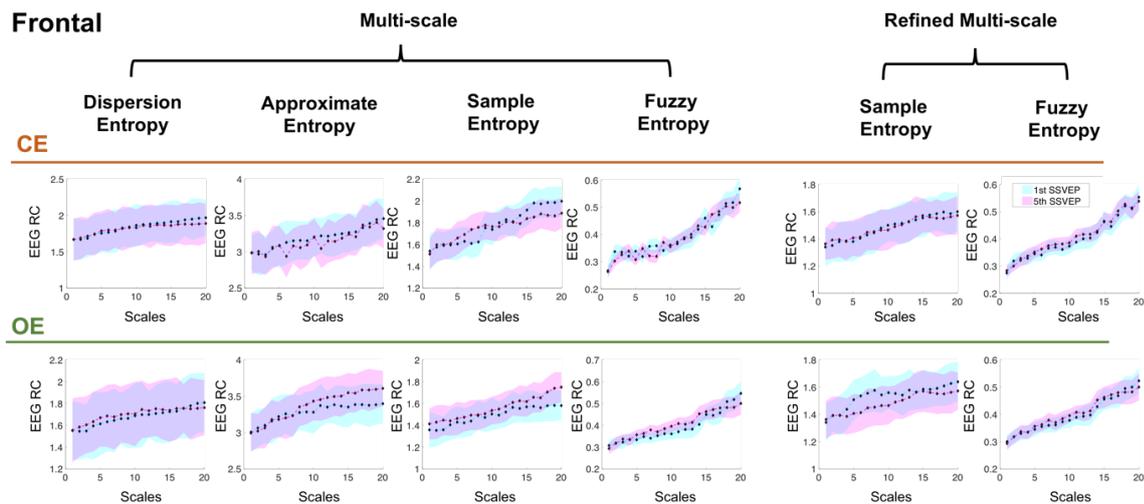

Figure 6 The trends of the EEG RC in the frontal area. Comparison of the EEG RC between the first and fifth stimuli over time scales from 1 to 20 in the CE or OE condition. Of note, the cyan and red traces represent the mean SD of the EEG RC in the first and fifth stimuli session, respectively. The purple trace represents the overlap area of the SD of the EEG RC between the first and fifth stimuli session.

## 4. Discussion

Despite progress in visual stimulus research in recent years, extracting complex brain dynamics that can be used to determine the habituation of brain systems remains challenging. Our study made a discovery by investigating the changes in EEG complexity elicited by SSVEP stimuli by applying a multiscale inherent fuzzy entropy algorithm in CE and OE conditions. We found that occipital and prefrontal EEG complexity increased with increasing stimulus times in the OE condition. However, only occipital EEG complexity increased in the CE condition. Additionally, changes in occipital EEG complexity resulted in higher entropy values in the OE condition than in the CE condition.

Systemic analyses indicate that entropy dynamics reflect the nonlinear complex characteristics of the brain rather than the linear characteristics that allow it to adapt to constantly changing stimulus situations [28]. The associated multiscale entropy methods, which can be used to quantify brain complexity to express the robustness of brain systems, are crucial for quantifying the critical characteristics of nonlinear neuro-dynamics [29]. Diseased systems usually show lower entropy values than do healthy systems [13, 23]. Additionally, as neurons function on a millisecond time scale, biological systems functioning on different time scales that may exhibit different behaviors (e.g., nonstationarity, nonlinearity, and sensitivity) must be investigated. This approach, which is known as multiscale entropy, is currently widely used to study the full spectrum of time-dependent variations in the complexity of physiological systems.

Several studies have suggested that photic driving responses (e.g., SSVEPs) have a more diffuse character on cortical EEG and are thus not limited to the visual cortex. Rather, these responses are associated with the activation of a distributed network of cortical regions, including the lateral prefrontal cortex and the hippocampal formation [1]. Our study revealed that EEG complexity over the occipital and prefrontal regions, as measured using multiscale inherent fuzzy entropy, was altered by repetitive visual stimuli in CE and OE conditions. These monotonous alterations in EEG complexity were identified by the Oz and Fpz electrodes. Our results show that EEG complexity was enhanced in the occipital and prefrontal areas after repetitive visual stimulation, a result consistent with those of a previous study showing that EEG entropy displayed an increasing trend in response to long-term audio-visual stimulation [30]. Importantly, decreased complexity may represent reduced robustness of the brain system, while elevated complexity is strongly correlated with stable and accurate behavioral performance [31]. Our finding that EEG complexity increases with increasing stimulus times may reflect a strong ability to tolerate perturbations, which may be related to the functional or structural modification of brain systems.

On the other hand, humans learn to stop responding to a stimulus that is no longer biologically relevant [9, 10]. For example, humans may habituate to repeated visual stimulus when they learn that these stimuli have no effects. This habituation performance is a form of adaptive behavior, which is beneficial to reflecting the robustness of brain systems. Our finding that EEG complexity was enhanced upon repetitive SSVEPs might be another expression pattern of habituation.

We also observed that these complexity changes were not equivalent between the CE and OE conditions. Both conditions were characterized by increases in occipital complexity with increasing stimulus times, and the OE condition showed greater complexity than did the CE condition. However, the above trend was not easily discernable in the prefrontal area. It is possible that the activation and sensitivity of cortices involved in various electrophysiological properties, e.g., power spectra [32] and regional connectivity [33], vary; however, this idea is not completely understood. Additionally, a benefit of the CE condition SSVEP brain–computer interface is that it allows users to express intentions without needing to open their eyes, which is beneficial for some groups of patients, e.g., individuals with impaired oculomotor function [34]. Another benefit of the CE condition SSVEP experiment is that it minimizes the negative influence of migraine on patients, as these patients are sensitive to light, and migraine attacks can be triggered or worsened by light stimulation [35].

Due to the broad availability and cost-effectiveness of EEG, this method has been considered a noninvasive means of assessing dynamic changes in brain electrical activity. The rapid development of dry sensors and wearable devices [25, 36, 37] has led to a reduction in the preparatory work required for long-term monitoring. Moreover, the headband design with two electrodes (Oz and Fpz) is convenient for long-term monitoring and daily use [25, 36, 37]. Taking into consideration artifacts removal methods and high-frequency oscillations [38, 39], it is possible to implement EEG-based models in laboratories and real-world settings. Thus, the complexity characteristics of easily assessed brain regions, the wearable EEG solution, and the multiscale inherent fuzzy entropy algorithm make occipital and prefrontal EEG complexity an obtainable signature of brain–computer interface systems used for pattern recognition.

Some limitations of our study need to be considered. First, most of our participants were female university students. Thus, the results of this study may have been affected by gender bias. Moreover, as stated above, the study did not include middle-aged or elderly participants. Second, the placement of the EEG electrodes (Oz and Fpz sites) was limited by the design of the headband. Previous studies have demonstrated that the occipital and prefrontal regions respond to visual stimuli, but further studies using equipment with multiple channels will be helpful for investigating brain complexity in other brain regions. Third, the current study focused on a single flicker frequency, so a larger study covering multiple flicker frequencies is recommended to obtain a comprehensive, full-spectrum understanding of changes in brain complexity in a visual stimulus environment.

## 5. Conclusions

This large-scale study collected EEG signals to investigate changes in brain complexity elicited by repetitive SSVEP stimuli using the multiscale inherent fuzzy entropy algorithm with relative measurements. Entropy is a measurement of complex brain dynamics in multiple time scales in a visual stimulus environment. Our results highlighted the feasibility of using multiscale inherent fuzzy entropy measurements to compare EEG complexity between different conditions and stimulus sessions. Our findings showed that occipital and frontal EEG RC increases with increasing visual stimulus times in the OE condition, yet this trend was only observed in the occipital area in the CE condition. Furthermore, the performance of the multiscale inherent fuzzy entropy algorithm was superior to that of competing multiscale-based entropy methods in the repetitive visual task. The findings also showed that RC in the brain provides insight regarding the robustness of brain systems. Wearable EEG devices are a promising tool enabling EEG-based brain–computer interface systems to recognize the effects of repetitive SSVEPs on EEG complexity.

**Acknowledgements**

This work was supported in part by the Australian Research Council (ARC) under discovery grant DP180100670 and DP180100656. The research was also sponsored in part by the Army Research Laboratory and was accomplished under Cooperative Agreement Numbers W911NF-10-2- 0022 and W911NF-10-D-0002/TO 0023. The views and conclusions contained in this document are those of the authors and should not be interpreted as representing the official policies, either expressed or implied, of the Army Research Laboratory or the U.S. government. Furthermore, this work was partially supported by the Natural Science Foundation of Jiangsu Province under Grant BK20151274. Additionally, the authors would like to express their sincere appreciation to the anonymous reviewers for their insightful comments, which greatly improved the quality of this manuscript.

**Conflicts of Interest**
None.

**References**
[1] P. Sehatpour, S. Molholm, T.H. Schwartz, J.R. Mahoney, A.D. Mehta, D.C. Javitt, P.K. Stanton, J.J. Foxe, A human intracranial study of long-range oscillatory coherence across a frontal–occipital–hippocampal brain network during visual object processing, Proceedings of the National Academy of Sciences, 105 (2008) 4399-4404.
[2] H.-Y.J. Tang, B. Riegel, S.M. McCurry, M.V. Vitiello, Open-Loop Audio-Visual Stimulation (AVS): A Useful Tool for Management of Insomnia?, Applied psychophysiology and biofeedback, 41 (2016) 39-46.
[3] L.B.I. Shah, P. Klainin-Yobas, S. Torres, P. Kannusamy, Efficacy of psychoeducation and relaxation interventions on stress-related variables in people with mental disorders: a literature review, Archives of psychiatric nursing, 28 (2014) 94-101.
[4] A.M. Norcia, L.G. Appelbaum, J.M. Ales, B.R. Cottereau, B. Rossion, The steady-state visual evoked potential in vision research: a review, Journal of vision, 15 (2015) 4-4.
[5] M.A. Pastor, J. Artieda, J. Arbizu, M. Valencia, J.C. Masdeu, Human cerebral activation during steady-state visual-evoked responses, Journal of neuroscience, 23 (2003) 11621-11627.
[6] X. Zhao, Y. Chu, J. Han, Z. Zhang, SSVEP-based brain–computer interface controlled functional electrical stimulation system for upper extremity rehabilitation, IEEE Transactions on Systems, Man, and Cybernetics: Systems, 46 (2016) 947-956.
[7] S. Qiu, Z. Li, W. He, L. Zhang, C. Yang, C. Su, Teleoperation control of an exoskeleton robot using brain machine interface and visual compressive sensing, IEEE Transactions on Fuzzy Systems, (2016).
[8] P.A. Herman, G. Prasad, T.M. McGinnity, Designing an Interval Type-2 Fuzzy Logic System for Handling Uncertainty Effects in Brain-Computer Interface Classification of Motor Imagery Induced EEG Patterns, IEEE Transactions on Fuzzy Systems, (2016).
[9] P.M. Groves, R.F. Thompson, Habituation: a dual-process theory, Psychological review, 77 (1970) 419.
[10] R.F. Thompson, W.A. Spencer, Habituation: a model phenomenon for the study of neuronal substrates of behavior, Psychological review, 73 (1966) 16.

[11] G. Coppola, F. Pierelli, J. Schoenen, Habituation and migraine, Neurobiology of learning and memory, 92 (2009) 249-259.
[12] C.H. Rankin, T. Abrams, R.J. Barry, S. Bhatnagar, D.F. Clayton, J. Colombo, G. Coppola, M.A. Geyer, D.L. Glanzman, S. Marsland, Habituation revisited: an updated and revised description of the behavioral characteristics of habituation, Neurobiology of learning and memory, 92 (2009) 135-138.
[13] T. Takahashi, R.Y. Cho, T. Mizuno, M. Kikuchi, T. Murata, K. Takahashi, Y. Wada, Antipsychotics reverse abnormal EEG complexity in drug-naive schizophrenia: a multiscale entropy analysis, Neuroimage, 51 (2010) 173-182.
[14] J. Gao, J. Hu, F. Liu, Y. Cao, Multiscale entropy analysis of biological signals: a fundamental bi-scaling law, Frontiers in computational neuroscience, 9 (2015).
[15] S.M. Pincus, Approximate entropy as a measure of system complexity, Proceedings of the National Academy of Sciences, 88 (1991) 2297-2301.
[16] M. Rostaghi, H. Azami, Dispersion entropy: A measure for time-series analysis, IEEE Signal Processing Letters, 23 (2016) 610-614.
[17] J.S. Richman, J.R. Moorman, Physiological time-series analysis using approximate entropy and sample entropy, American Journal of Physiology-Heart and Circulatory Physiology, 278 (2000) H2039-H2049.
[18] J.F. Valencia, A. Porta, M. Vallverdu, F. Claria, R. Baranowski, E. Orlowska-Baranowska, P. Caminal, Refined multiscale entropy: Application to 24-h holter recordings of heart period variability in healthy and aortic stenosis subjects, IEEE Transactions on Biomedical Engineering, 56 (2009) 2202-2213.
[19] W. Chen, Z. Wang, H. Xie, W. Yu, Characterization of surface EMG signal based on fuzzy entropy, IEEE Transactions on neural systems and rehabilitation engineering, 15 (2007) 266-272.
[20] L. Ji, P. Li, K. Li, X. Wang, C. Liu, Analysis of short-term heart rate and diastolic period variability using a refined fuzzy entropy method, Biomedical engineering online, 14 (2015) 64.
[21] M. Costa, A.L. Goldberger, C.-K. Peng, Multiscale entropy analysis of biological signals, Physical review E, 71 (2005) 021906.
[22] Z. Cao, C.-T. Lin, Inherent Fuzzy Entropy for the Improvement of EEG Complexity Evaluation, IEEE Transactions on Fuzzy Systems, (2017).
[23] Z. Cao, K.-L. Lai, C.-T. Lin, C.-H. Chuang, C.-C. Chou, S.-J. Wang, Exploring resting-state EEG complexity before migraine attacks, Cephalalgia, (2017) 0333102417733953.
[24] M.E. Bouton, Learning and behavior: A contemporary synthesis (Sinauer Associates, 2007).
[25] C.-T. Lin, L.-D. Liao, Y.-H. Liu, I.-J. Wang, B.-S. Lin, J.-Y. Chang, Novel dry polymer foam electrodes for long-term EEG measurement, Biomedical Engineering, IEEE Transactions on, 58 (2011) 1200-1207.
[26] R. Kuś, A. Duszyk, P. Milanowski, M. Łabęcki, M. Bierzyńska, Z. Radzikowska, M. Michalska, J. Żygierewicz, P. Suffczyński, P.J. Durka, On the quantification of SSVEP frequency responses in human EEG in realistic BCI conditions, PloS one, 8 (2013) e77536.
[27] N.E. Huang, Z. Shen, S.R. Long, M.C. Wu, H.H. Shih, Q. Zheng, N.-C. Yen, C.C. Tung, H.H. Liu, The empirical mode decomposition and the Hilbert spectrum for nonlinear and non-stationary time series analysis, Proceedings of the Royal Society of London A: mathematical, physical and engineering sciences, (The Royal Society1998), pp. 903-995.
[28] M. Gazzaniga, K. Doron, C. Funk, Looking toward the future: Perspectives on examining the architecture and function of the human brain as a complex system, The cognitive neurosciences IV, (2010) 1245-1252.
[29] J. Gao, J. Hu, W.-w. Tung, Complexity measures of brain wave dynamics, Cognitive neurodynamics, 5 (2011) 171-182.
[30] M. Teplan, A. Krakovska, S. Štolc, EEG responses to long-term audio–visual stimulation, International journal of psychophysiology, 59 (2006) 81-90.

[31] S. Lippé, N. Kovacevic, A.R. McIntosh, Differential maturation of brain signal complexity in the human auditory and visual system, Frontiers in Human Neuroscience, 3 (2009).
[32] R.J. Barry, A.R. Clarke, S.J. Johnstone, C.A. Magee, J.A. Rushby, EEG differences between eyes-closed and eyes-open resting conditions, Clinical Neurophysiology, 118 (2007) 2765-2773.
[33] Z. Cao, C.-T. Lin, C.-H. Chuang, K.-L. Lai, A.C. Yang, J.-L. Fuh, S.-J. Wang, Resting-state EEG power and coherence vary between migraine phases, The journal of headache and pain, 17 (2016) 102.
[34] J.-H. Lim, H.-J. Hwang, C.-H. Han, K.-Y. Jung, C.-H. Im, Classification of binary intentions for individuals with impaired oculomotor function:'eyes-closed'SSVEP-based brain–computer interface (BCI), Journal of neural engineering, 10 (2013) 026021.
[35] G. Strigaro, A. Cerino, L. Falletta, D. Mittino, C. Comi, C. Varrasi, R. Cantello, Impaired visual inhibition in migraine with aura, Clinical Neurophysiology, 126 (2015) 1988-1993.
[36] B.C.-T. Lin, L.-W. Ko, J.-C. Chiou, J.-R. Duann, R.-S. Huang, S.-F. Liang, T.-W. Chiu, T.-P. Jung, Noninvasive neural prostheses using mobile and wireless EEG, Proceedings of the IEEE, 96 (2008) 1167-1183.
[37] L.-D. Liao, C.-T. Lin, K. McDowell, A.E. Wickenden, K. Gramann, T.-P. Jung, L.-W. Ko, J.-Y. Chang, Biosensor technologies for augmented brain–computer interfaces in the next decades, Proceedings of the IEEE, 100 (2012) 1553-1566.
[38] J. Hu, C.-s. Wang, M. Wu, Y.-x. Du, Y. He, J. She, Removal of EOG and EMG artifacts from EEG using combination of functional link neural network and adaptive neural fuzzy inference system, Neurocomputing, 151 (2015) 278-287.
[39] M. Wu, T. Wan, X. Wan, Y. Du, J. She, Fast, accurate localization of epileptic seizure onset zones based on detection of high-frequency oscillations using improved wavelet transform and matching pursuit methods, Neural computation, 29 (2017) 194-219.